\documentclass[sigconf]{acmart}
\DeclareUnicodeCharacter{2248}{\ensuremath{\approx}}
\AtBeginDocument{%
  }

\usepackage{multirow}   
\usepackage{array} 
\usepackage{graphicx}
\usepackage{subfigure}
\usepackage{subcaption}
\usepackage{caption}
\usepackage[font=small, labelfont=bf]{caption}
\begin{document}

\title{Rethinking Temporal Models for TinyML: LSTM versus 1D-CNN in Resource-Constrained Devices}

\author{Bidyut Saha}
\affiliation{%
  \institution{Sister Nivedita University, Newtown} 
  \country{India}}
\email{bidyut.s@snuniv.ac.in}

\author{Riya Samanta}
\affiliation{%
  \institution{Techno India University, Salt Lake}
  \country{India}}
\email{riya.s@technoindiaeducation.com}

\renewcommand{\shortauthors}{Bidyut et al.}

\begin{abstract}

Time series classification underpins applications such as human activity recognition, healthcare monitoring, and gesture detection in the IoT domain. Tiny Machine Learning (TinyML) enables models to run directly on low-power microcontroller units (MCUs), improving efficiency, ensuring privacy, and reducing cost by avoiding reliance on cloud or edge computing. While Long Short-Term Memory (LSTM) networks are widely used for capturing temporal dependencies, their high computational and memory demands make real-time MCU deployment impractical. In this work, we conduct a hardware-aware feasibility study of LSTM versus 1D-Convolutional Neural Networks (1D-CNN) across five benchmark datasets. Results show that 1D-CNN consistently achieves comparable or higher accuracy (≈95\%) than LSTM (≈89\%), while requiring 35\% less RAM, approx. 25\% less Flash, and enabling real-time inference (27.6 ms vs. 2038 ms). Being so lightweight, 1D-CNN is particularly suitable for on-device processing in wearables and other low-power, battery-operated systems, establishing it as a practical and resource-efficient choice for TinyML deployment.

\end{abstract}

\ccsdesc[300]{Human-centered computing~Ubiquitous and mobile computing}
\ccsdesc[300]{Computing methodologies~Supervised learning by classification}
\ccsdesc[300]{Hardware~Emerging architectures}
\ccsdesc[300]{Computer systems organization~Embedded systems}

\keywords{ TinyML, Time Series Classification, 1D-CNN, LSTM, Resource-Constrained Devices, Microcontroller Units, On-Device Machine Learning, Human Activity Recognition, Energy-Efficient AI}

\maketitle

\section{Introduction}

Time-series classification powers applications such as human activity recognition, gesture detection, and healthcare monitoring. While Long Short-Term Memory (LSTM) networks are a standard choice for modeling temporal dependencies \cite{hochreiter1997long}, their high memory and computational costs make them impractical for real-time on-device computting on microcontroller units (MCUs), which typically provide only a few hundred kilobytes of SRAM and less than 1 MB of flash. Existing solutions often rely on edge or cloud offloading \cite{satyanarayanan2017emergence}, but this increases latency, energy consumption, infrastructure cost \cite{varghese2018next},  risks to data privacy \cite{zhao2017time} and connectivity dependency \cite{shi2016edge}.

Tiny Machine Learning (TinyML) enables on-device inference directly on low-cost, battery-operated MCUs, offering efficiency, privacy, and affordability \cite{banbury2021micronets}. However, to be viable, models must balance accuracy with strict resource constraints. One promising alternative is the 1D-Convolutional Neural Network (1D-CNN), which, though originally designed for spatial features, has shown potential for time-series tasks with significantly lower resource footprints.


\noindent \textbf{Contributions of this work are:}
\begin{enumerate}

    \item A hardware-aware, apples-to-apples comparison of LSTM and 1D-CNN architectures for TinyML time-series classification.
    
    \item A deployability-focused evaluation protocol covering accuracy, RAM/Flash usage, inference latency, and quantization effects on MCUs.

    \item Empirical evidence from five benchmark datasets across human–computer interaction, lifestyle, and healthcare domains, showing that 1D-CNNs consistently achieve comparable or higher accuracy than LSTMs while requiring fewer resources and supporting real-time performance.

\end{enumerate}

\noindent These results demonstrate that 1D-CNNs are lightweight and practical for on-device processing, particularly in wearable and low-power, battery-operated systems.


\vspace{-0.1in}
\begin{figure*}
    \centering
    \subfigure[]{\includegraphics[scale=0.25]{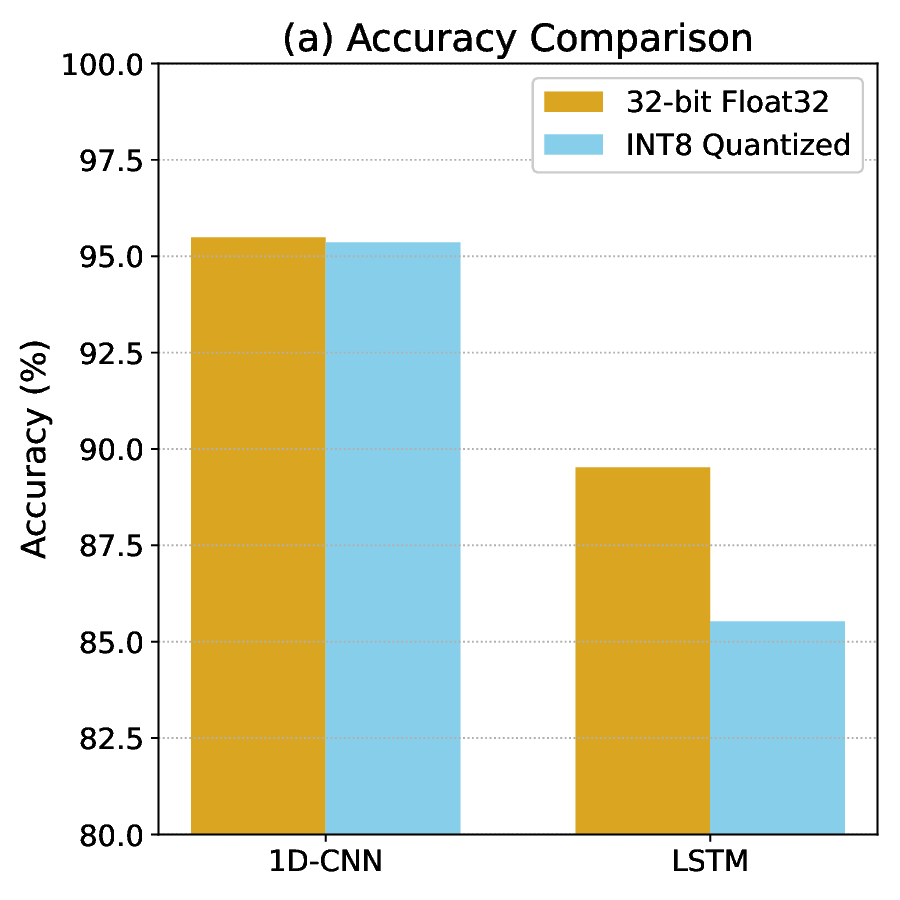}\label{Fig3a}}
    \subfigure[]{\includegraphics[scale=0.25]{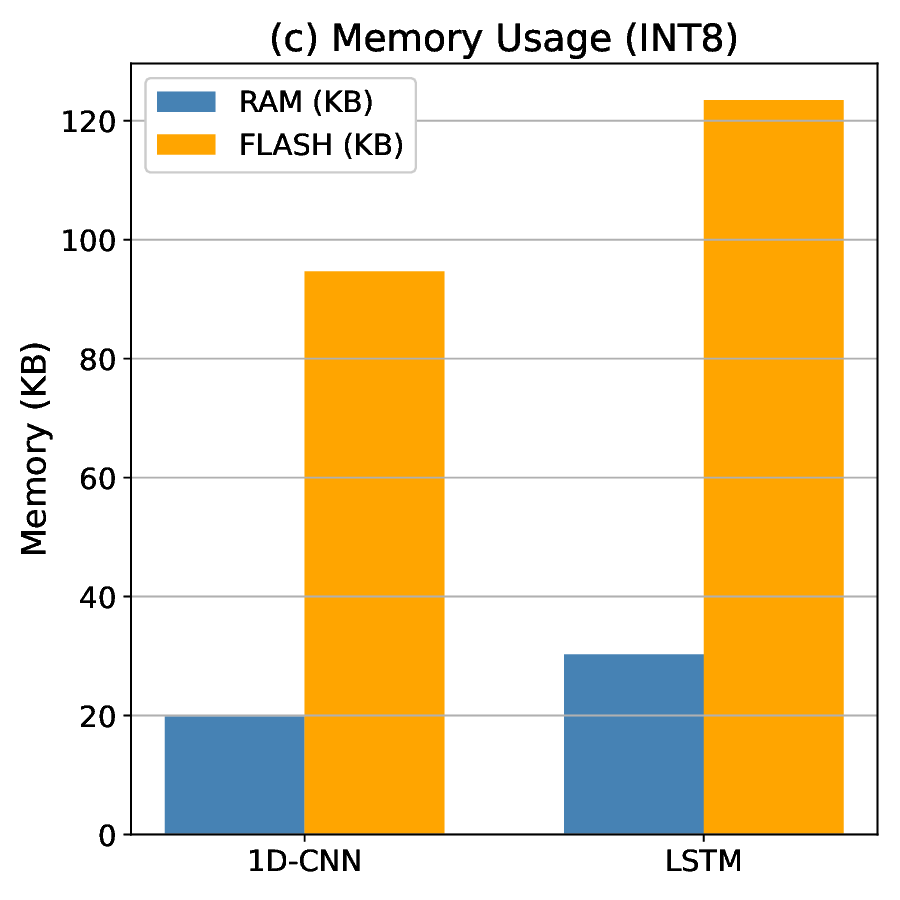}\label{Fig3b}}
    \subfigure[]{\includegraphics[scale=0.25]{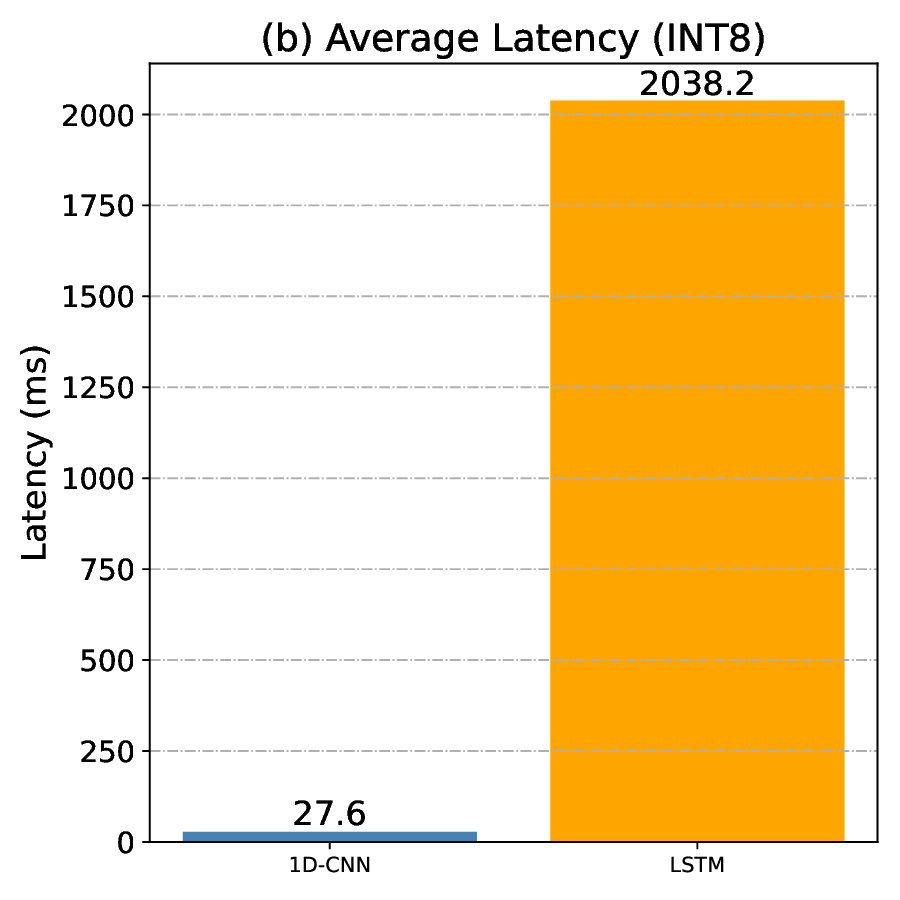}\label{Fig3c}}

   \vspace{-0.2in}
    
    \caption{Average performance of 1D-CNN \& LSTM across 5 datasets on ESP32: (a) accuracy , (b) memory usage, and (c)  inference latency (ms).}
    \label{Fig3}
    \vspace{-0.1in}
\end{figure*}


\begin{table}[!t]
\centering
\footnotesize
\caption{Performance of 1D-CNN vs. LSTM on five benchmark datasets under Float32 and INT8 quantization, reporting accuracy, RAM, FLASH, and inference latency on ESP32.}
\vspace{-0.1in}
\label{tab:cnn_lstm_results}
\begin{tabular}{c|c|c|c|c|c}
\hline
\textbf{Dataset} & \textbf{Model} & \textbf{Acc. (32/INT8)} & \textbf{RAM} & \textbf{FLASH} & \textbf{Latency} \\
                 &                & (\%)                    & (KB)         & (KB)           & (ms)             \\
\hline
\multirow{2}{*}{UCIHAR} 
 & 1D-CNN & 93.34 / 92.03 & 20.38 & 95.12  & 28    \\
 & LSTM   & 92.38 / 88.09 & 28.01 & 124.02 & 2356  \\
\hline
\multirow{2}{*}{PAMAP2} 
 & 1D-CNN & 96.12 / 95.60 & 13.59 & 94.67  & 10    \\
 & LSTM   & 95.81 / 90.80 & 15.27 & 123.66 & 746   \\
\hline
\multirow{2}{*}{WISDM} 
 & 1D-CNN & 94.77 / 94.34 & 13.59 & 94.92  & 20    \\
 & LSTM   & 93.55 / 89.21 & 15.27 & 123.65 & 502   \\
\hline
\multirow{2}{*}{MITBIH} 
 & 1D-CNN & 97.24 / 97.15 & 24.84 & 94.28  & 40    \\
 & LSTM   & 85.21 / 78.22 & 38.47 & 123.25 & 3233  \\
\hline
\multirow{2}{*}{PTB} 
 & 1D-CNN & 97.97 / 97.66 & 24.84 & 94.31  & 40    \\
 & LSTM   & 82.65 / 79.35 & 36.45 & 122.71 & 3354  \\
\hline
\multirow{2}{*}{Average} 
 & 1D-CNN & \textbf{95.49 / 95.36} & \textbf{19.85} & \textbf{94.66}  & \textbf{27.6}   \\
 & LSTM   & \textbf{89.52 / 85.53} & \textbf{30.29} & \textbf{123.46} & \textbf{2038.2} \\
\hline
\end{tabular}
\vspace{-0.1in}
\end{table}


\section{Methodology}




\paragraph{Datasets \& Preprocessing.}
We evaluate on five benchmark datasets. From the \textit{human-computer interaction and lifestyle} domain, we use UCI-HAR \cite{misc_human_activity_recognition_using_smartphones_240}, PAMAP2 \cite{misc_pamap2_physical_activity_monitoring_231}, and WISDM \cite{misc_wisdm_smartphone_and_smartwatch_activity_and_biometrics_dataset__507}, which contain IMU signals for human activity recognition. From the \textit{healthcare} domain, we include the MIT-BIH Arrhythmia Database \cite{moody2001impact} and the PTB Diagnostic ECG Database \cite{bousseljot1995nutzung}, both of which provide ECG signals over time. Data preprocessing follows the pipeline described in \cite{saha2024tinytnas}.

\paragraph{Model Families.}

We select two representative architectures — a one-dimensional convolutional neural network (1D-CNN) and a long short-term memory network (LSTM) — as they exemplify the two primary modeling approaches for time-series data: local pattern extraction through convolutional operations (1D-CNN) and recurrent modeling of long-range temporal dependencies (LSTM). For clarity, the 1D-CNN used throughout this work refers specifically to a depthwise separable convolutional implementation, which provides computational efficiency while maintaining representational capacity. Rather than exhaustively evaluating a wide range of architectural variants, we focus on canonical, capacity-controlled versions of these models. We further conduct targeted ablation studies  to demonstrate that the observed trade-offs generalize across typical design choices.

For our experiments, the two architectures are implemented as follows:

\begin{enumerate}
    \item \textbf{1D-CNN Model:} Three separable convolutional layers (32, 48, and 72 filters, kernel size $=3$), each followed by max pooling. A global average pooling layer reduces temporal dimension, followed by a dense layer (72 units, ReLU), dropout, and a final softmax output layer.  

    \item \textbf{LSTM Model:} A stacked LSTM with 32, 48, and 72 units to capture long-range temporal dependencies. The final output is passed through a dense layer (72 units, ReLU), dropout, and a softmax output layer.  
\end{enumerate}

\paragraph{Training Protocol.}
Both models use the same optimizer, batch size, and epoch budget, with early stopping on validation loss. Hyperparameters (learning rate, kernel sizes/hidden units) are selected via a small grid search using validation performance, avoiding test leakage. Random seeds are fixed for reproducibility.  

\paragraph{Metrics.}
We report classification accuracy for both the non-quantized Float32 models and the INT8-quantized counterparts as the primary performance measure. In addition, we evaluate model size (Flash, in kB), peak RAM usage during inference (in kB), and on-device latency (ms) measured on an ESP32 MCU operating at 240~MHz. Resource footprints are consistently reported for the INT8-quantized models.

\paragraph{Optimization and On-Device Execution.}
To ensure that the models are suitable for deployment on microcontroller units (MCUs) under strict resource constraints, we design lightweight versions of the architectures. Further optimization is achieved through quantization. Specifically, we apply INT8 post-training quantization using the \textit{TensorFlow Model Optimization Toolkit} to reduce the model size and computational requirements. For on-device execution, the optimized models are deployed using the \textit{TensorFlow Lite Micro} framework on the ESP32 MCU platform.

\section{Results and Discussion}

Table~\ref{tab:cnn_lstm_results} presents the detailed results for all five datasets using both 1D-CNN and LSTM models. For accuracy, we report both the non-quantized Float32 and the INT8-quantized versions. As shown in Figure~\ref{Fig3}(a), where the averaged accuracies are plotted, 1D-CNN consistently outperforms LSTM under both Float32 and INT8 settings. An important observation is that accuracy degradation due to INT8 quantization is negligible for 1D-CNN, whereas LSTM suffers a significant drop in performance, making CNN models more robust under quantization.  Figure~\ref{Fig3}(b) further illustrates the memory advantage: 1D-CNN requires substantially less RAM and Flash compared to LSTM. Likewise, Figure~\ref{Fig3}(c) highlights the extreme gap in latency—while 1D-CNN maintains real-time execution (27.6~ms on average), LSTM incurs prohibitive runtimes exceeding 2~seconds per inference.  

Overall, the combined tabular and graphical evidence suggests that in extremely resource-constrained TinyML environments—where INT8 quantization is often mandatory—1D-CNN offers a far more practical choice over LSTM, providing lower latency and smaller memory footprints while sustaining higher accuracy.

\vspace{-0.05in}
\section{Limitations and Future Work.}
This study provides a solid foundation for future exploration in forecasting and other time-series applications beyond classification. The feature shapes considered—approximately 2 seconds in duration with sampling rates up to 50~Hz—represent practical scenarios for many TinyML use cases. Building on these results, future work will extend experiments to longer sequences, higher-frequency data, and additional application domains, further enhancing the robustness and versatility of the proposed methods.

\bibliographystyle{ACM-Reference-Format}
\bibliography{sample-base}

@article{banbury2021micronets,
  title={Micronets: Neural network architectures for deploying tinyml applications on commodity microcontrollers},
  author={Banbury, Colby and Zhou, Chuteng and Fedorov, Igor and Matas, Ramon and Thakker, Urmish and Gope, Dibakar and Janapa Reddi, Vijay and Mattina, Matthew and Whatmough, Paul},
  journal={Proceedings of machine learning and systems},
  volume={3},
  pages={517--532},
  year={2021}
}

@article{hochreiter1997long,
  title={Long short-term memory},
  author={Hochreiter, Sepp and Schmidhuber, J{\"u}rgen},
  journal={Neural computation},
  volume={9},
  number={8},
  pages={1735--1780},
  year={1997},
  publisher={MIT press}
}

@article{satyanarayanan2017emergence,
  title={The emergence of edge computing},
  author={Satyanarayanan, Mahadev},
  journal={Computer},
  volume={50},
  number={1},
  pages={30--39},
  year={2017},
  publisher={IEEE}
}

@article{shi2016edge,
  title={Edge computing: Vision and challenges},
  author={Shi, Weisong and Cao, Jie and Zhang, Quan and Li, Youhuizi and Xu, Lanyu},
  journal={IEEE internet of things journal},
  volume={3},
  number={5},
  pages={637--646},
  year={2016},
  publisher={Ieee}
}

@article{varghese2018next,
  title={Next generation cloud computing: New trends and research directions},
  author={Varghese, Blesson and Buyya, Rajkumar},
  journal={Future Generation Computer Systems},
  volume={79},
  pages={849--861},
  year={2018},
  publisher={Elsevier}
}

@inproceedings{zhao2017time,
  title={Time-weighted LSTM model with redefined labeling for stock trend prediction},
  author={Zhao, Zhiyong and Rao, Ruonan and Tu, Shaoxiong and Shi, Jun},
  booktitle={2017 IEEE 29th international conference on tools with artificial intelligence (ICTAI)},
  pages={1210--1217},
  year={2017},
  organization={IEEE}
}

@misc{misc_human_activity_recognition_using_smartphones_240,
  author       = {Reyes Ortiz et. al},
  title        = {{Human Activity Recognition Using Smartphones}},
  year         = {2012},
  howpublished = {UCI Machine Learning Repository},
  note         = {{DOI}: https://doi.org/10.24432/C54S4K}
}

@misc{misc_pamap2_physical_activity_monitoring_231,
  author       = {Reiss,Attila},
  title        = {{PAMAP2 Physical Activity Monitoring}},
  year         = {2012},
  howpublished = {UCI Machine Learning Repository},
  note         = {{DOI}: https://doi.org/10.24432/C5NW2H}
}

@misc{misc_wisdm_smartphone_and_smartwatch_activity_and_biometrics_dataset__507,
  author       = {Weiss,Gary},
  title        = {{WISDM Smartphone and Smartwatch Activity and Biometrics Dataset }},
  year         = {2019},
  howpublished = {UCI Machine Learning Repository},
  note         = {{DOI}: https://doi.org/10.24432/C5HK59}
}

@article{moody2001impact,
  title={The impact of the MIT-BIH arrhythmia database},
  author={Moody, George B and Mark, Roger G},
  journal={IEEE engineering in medicine and biology magazine},
  volume={20},
  number={3},
  pages={45--50},
  year={2001},
  publisher={IEEE}
}

@article{bousseljot1995nutzung,
  title={Nutzung der EKG-Signaldatenbank CARDIODAT der PTB {\"u}ber das Internet},
  author={Bousseljot, Ralf and Kreiseler, Dieter and Schnabel, Allard},
  year={1995},
  publisher={Walter de Gruyter, Berlin/New York Berlin, New York}
}

@article{saha2024tinytnas,
  title={Tinytnas: Gpu-free, time-bound, hardware-aware neural architecture search for tinyml time series classification},
  author={Saha, Bidyut and Samanta, Riya and Ghosh, Soumya K and Roy, Ram Babu},
  journal={arXiv preprint arXiv:2408.16535},
  year={2024}
}

\end{document}